\documentclass[preprint,12pt]{article}

\usepackage{amsmath}
\usepackage{graphicx}
\usepackage{pdfpages}
\usepackage[a4paper, margin=1.25in]{geometry}
\usepackage{authblk}
\usepackage{url}
\usepackage{cite}
\usepackage{csquotes}
\usepackage{float}
\usepackage{algorithm2e}

\title{Simple and accurate complete elliptic integrals for the full range of modulus}
\author{Teepanis Chachiyo  \\ email: teepanisc@nu.ac.th}
\affil{Department of Physics, Faculty of Science, \\ Naresuan University, Phitsanulok 65000, Thailand}
\date{\today}

\begin{document}
\maketitle

\begin{abstract}
The complete elliptic integral of the first and second kind, K(k) and E(k), appear in a multitude of physics and engineering applications. Because there is no known closed-form, the exact values have to be computed numerically. Here, approximations for the integrals are proposed based on their asymptotic behaviors.  An inverse of K is also presented. As a result, the proposed K(k) and E(k) reproduce the exact analytical forms both in the zero and asymptotic limits, while in the mid-range of modulus maintain average error of 0.06\% and 0.01\% respectively.  The key finding  is the ability to compute the integrals with exceptional accuracy on both limits of elliptical conditions. An accuracy of 1 in 1,000  should be sufficient for practical or prototyping engineering and architecture designs. The simplicity should facilitate discussions of advanced physics topics in introductory physics classes, and enable broader collaborations among researchers from other fields of expertise. For example, the phase space of energy-conserving nonlinear pendulum using only elementary functions is discussed. The proposed inverse of K is shown to be Never Failing Newton Initialization and is an important step for the computation of the exact inverse. An algorithm based on Arithmetic-Geometric Mean for computing exact integrals and their derivatives are also presented, which should be useful in a platform that special functions are not accessible such as web-based and firmware developments. Comparisons with sharp bounds from the mathematical inequalities literature further highlight the competitiveness of the proposed approximations.
\end{abstract}

\medskip
Keywords: computational physics, nonlinear pendulum,  complete elliptic integral, mathematical inequality

\section{Introduction}

Complete elliptic integrals are important mathematical tools for physics and engineering \cite{Marion1995}. Formally, there are three kinds of the integrals, but only two of which  are more frequently used.

The complete elliptic integral of the first kind, denoted $K(k)$, is defined as:

\begin{equation}
	K(k) = \int_{0}^\frac{\pi}{2} \frac{d\theta}{\sqrt{1-k^2 \sin^2 \theta}},
\end{equation}
and that of the second kind:

\begin{equation}
	E(k) = \int_{0}^\frac{\pi}{2} \sqrt{1-k^2 \sin^2 \theta} \, d\theta.
\end{equation}

\begin{figure}[h]
	\begin{center}
		\includegraphics[width=0.8\textwidth]{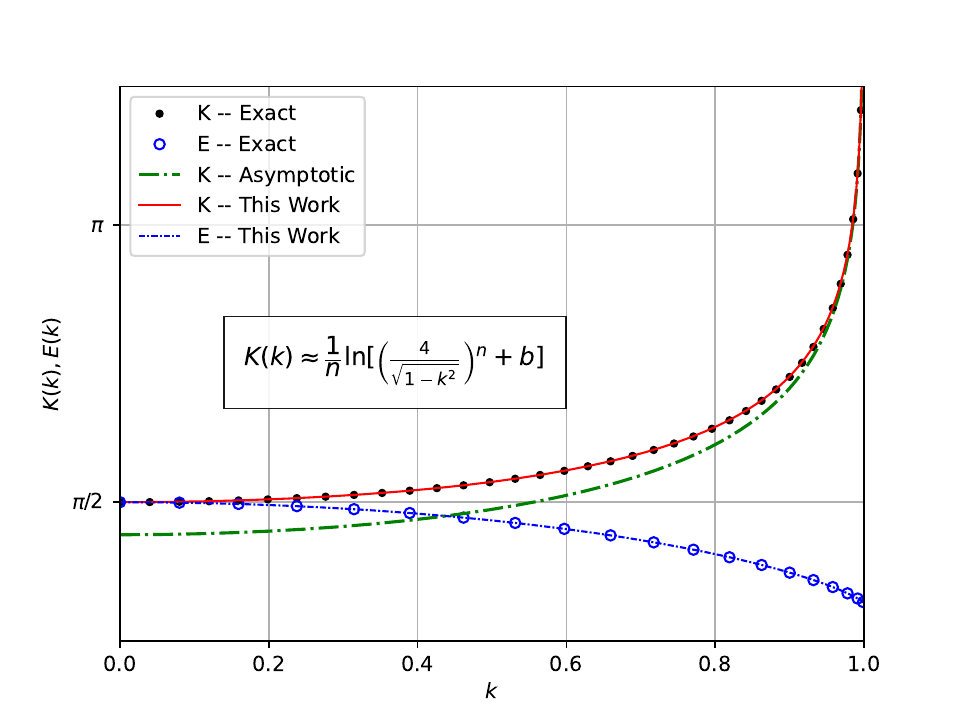}
		\caption{\label{fig_KE} Exact and well-known approximations of $K(k)$ and $E(k)$}
	\end{center}
\end{figure}

The argument $k$ is called \textquote{modulus} which is defined inside the range $[0,1]$. The exact values for the two integrals are shown in Fig. \ref{fig_KE}. They start at the same point $\pi/2$ and then branch off into separate directions. One continues to increase, approaching infinity; and the other decreases, approaching one.

The two integrals appear in a multitude of  physics and engineering applications. For example, $K(k)$ is needed to compute the period of a nonlinear pendulum \cite{Marion1995}. A nonlinear pendulum itself is being explored for applications in alternative energy sources \cite{Feng2024, Fu2023, Marszal2017} and artificial intelligence \cite{Krishnapriyan2023}. Therefore, the integral is needed in these emerging developments as well. $E(k)$ is needed to compute a circumference of an ellipse, a basic shape used in engineering such as architecture designs, satellite orbits, mechanical gears \cite{Bair2002}, and alternative energy sources \cite{Alom2016, Sutaji2021}. Specifically, an elliptical gear can provide variable speed in applications requiring non-uniform motions. An elliptical turbine has higher efficiency over its circular counterpart and has been used for harvesting energy from offshore wind  and from building drainage systems \cite{Sutaji2021}. Therefore, the complete elliptic integrals are one of the nuts and bolts of these research and developments.

$K(k)$ and $E(k)$ can be calculated exactly with computer programming via a numerical technique called Arithmetic-Geometric Mean (AGM) \cite[Eq.17.6.3-4]{Abramowitz1965}. However, the search for their approximate analytical form remains an active area of research \cite{Fukushima2013, Boyd2015, Yang2020, Zhu2022}. A wide variety of these approximations has been published \cite{Hinrichsen2021}, some of which are presented as pendulum period \cite{Ganley1985, Lima2008}, but is essentially an approximation of $K(k)$.

Although analytical approximations of $K(k)$ are already available, they are either very accurate but complex, not applicable for all $k$, or  simple  but much less accurate. A formula that is both simple and accurate is rare. Even rarer in the literature is a simple form of an inverse of $K(k)$ \cite{Boyd2015}.

An inverse of $K(k)$ in this sense is to find a map $K \rightarrow k$. For example, given a desired period of a pendulum, what would be its corresponding amplitude? It is true that an incomplete elliptic integral of the first kind $F(\phi, k)$, a more general form of $K(k)$, already has a well-known inverse directly related to  the Jacobi elliptic sine function $\text{sn}(u, k)$. As discussed by Boyd \cite{Boyd2015}, it is an inverse with respect to the angle $\phi$, not the modulus parameter $k$.  Therefore, $\text{sn}(u,k)$ does not help when trying to find  a map $K \rightarrow k$. As a result, the \emph{inverse} of $K(k)$ is not available in a standard mathematical software package.

This work aims to fill the research gaps listed above, and to strike a balance between simplicity and accuracy while maintaining the full range of applicability.

\section{Results}

In this report, an analytical formula for the complete elliptical integrals are proposed, offering significant improvements in simplicity and accuracy over existing ones. The first kind is of the form

\begin{equation}
	K(k) \approx  \frac{1}{ n } \ln [   \Big(\frac{4}{\sqrt{1-k^2}} \Big)^n + b ],
	\label{eq_K_thiswork}
\end{equation}

\noindent where the constant $n = \frac{\ln 4 -  \ln \pi}{\pi/2 - \ln 4}, \, b = e^{n\pi/2} - 4^n$.  Another function that is useful because it rarely is  available in a standard software package, is an inverse of $K(k)$. Namely,

\begin{equation}
k(K) \approx \sqrt{1-16/(e^{nK}-b)^{2/n}} \label{eq_invK_thiswork}.
\end{equation}

Using the same approximation technique, the complete elliptic integral of the second kind is

\begin{equation}
E(k)  \approx 1 + \frac{1-k^2}{2 n}\ln [   \Big(\frac{4/\!\sqrt{e}}{\sqrt{1-k^2}}\Big)^n + b  ]
\label{eq_E_thiswork},
\end{equation}

\noindent where the constant $ n = \frac{  \ln(3\pi/2 - 4)\,}{ \ln 4 - \pi + 3/2 } ,\, b = e^{n(\pi-2)} - (4/\!\sqrt{e})^n$. Note that the constants $\{n, b\}$ for the two types  are different.

\begin{figure}[ht]
	\begin{center}
		\includegraphics[width=0.8\textwidth]{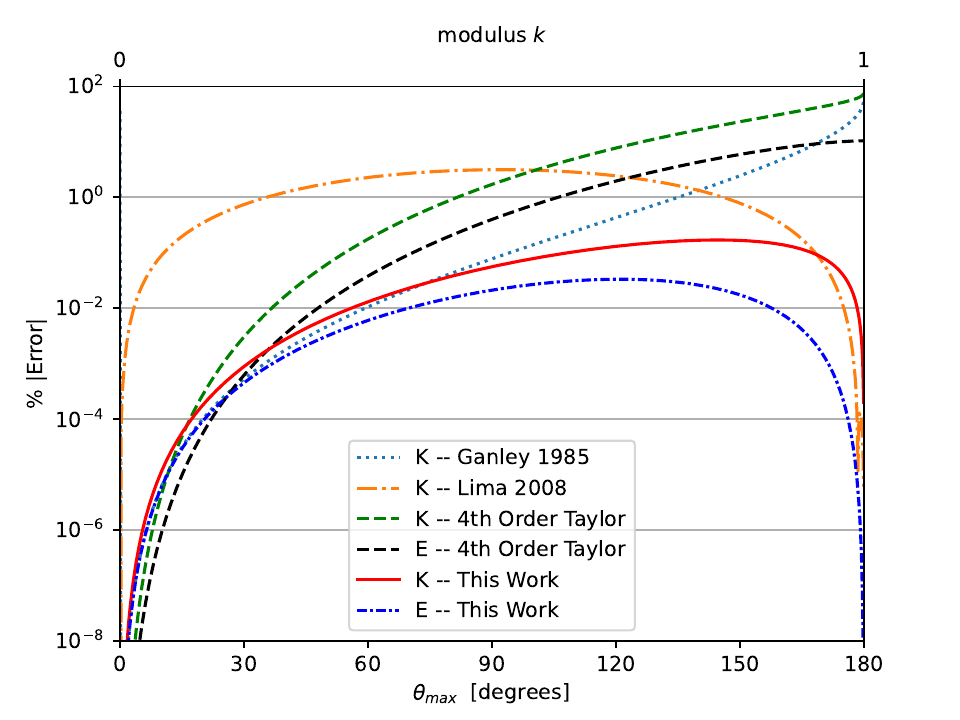}
		\caption{\label{fig_percent} Percent error of $K(k)$ and $E(k)$ from various simple analytical forms}
	\end{center}
\end{figure}

By design, the above formulae are exact when $k$ approaches the two limiting cases: $k \rightarrow 0$ and  $k \rightarrow 1$. Numerical evidence of the exactness at the two opposing limits is shown in Fig.~{\ref{fig_percent}}. 

Fig.~\ref{fig_percent} shows the percent error of various approximation methods. To aid direct applications in engineering, amplitudes of the pendulum $\theta_{max}$ (in degrees) are used as the horizontal axis instead of $k$.  The two parameters are related via $k = \sin(\theta_{max}/2)$. The work of Ganley  in 1985 represents early attempts to a find simple analytical form of $K(k)$ \cite{Ganley1985}; while the Lima formula is the first attempt to cover the full range of modulus.

The log-scale percent error clearly indicates that the errors drop quickly to zero on both sides. Therefore, the graph numerically proves that the formulae in this work are exact on both limits.

For the proposed formula $K(k)$, the average error is 0.06\% with 0.17\% maximum at the higher modulus range. The proposed formula $E(k)$ shares the same feature of the error profile, but much smaller quantitatively. The average is 0.01\% with only 0.03\% maximum.

\begin{figure}[ht]
	\begin{center}
		\includegraphics[clip,trim=50 0 50 0,width=1.0\textwidth]{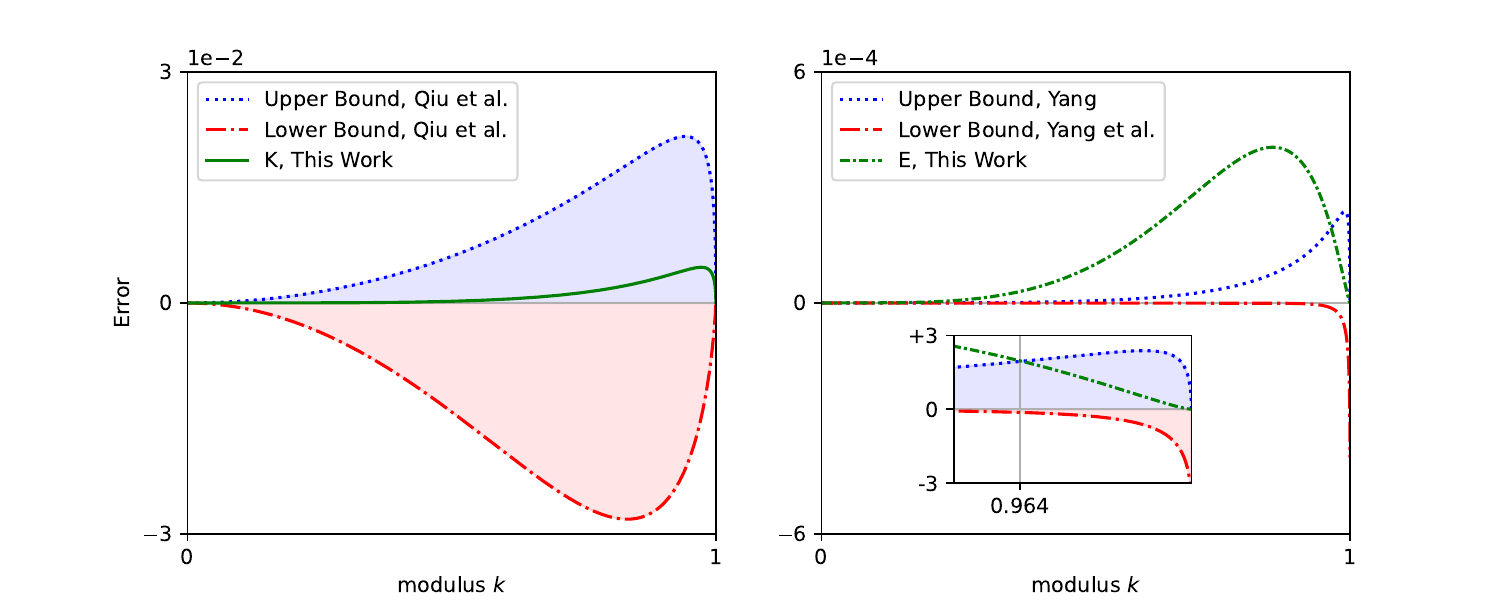}
		\caption{\label{fig_sharp_bound}  Error of $K(k)$ and $E(k)$ from various mathematical inequalities}
	\end{center}
\end{figure}

In addition, there exists a substantial body of work in the mathematical inequalities literature (e.g., \cite{Qiu1998_ra1, Alzer1998_ra2, Alzer2004_ra3, Yang2016_ra4, Yang2018_ra5, Yang2018_ra6, Yang2020_ra7, Yang2021_ra8}) that establishes sharp bounds and monotonicity results for the complete elliptic integrals. A bound (inequality) is a proposed mathematical expression that has been rigorously proved to be either higher or lower than the special function of interest, typically referred to as upper or lower bound. The bound is \emph{sharp} if its parameters have been optimized, relative to the proposed form of the inequality. Notable examples of simple and accurate sharp bounds are shown in Fig.~\ref{fig_sharp_bound}, together with the present approximations for comparison.

Fig.~\ref{fig_sharp_bound} (left) shows the performance of the upper and lower bounds for the complete elliptic integral $K(k)$. By resolving conjectures on inequalities between the $K(k)$ and simple logarithmic expressions, Qiu and coauthors derived sharp two-sided bounds for $K(k)$ \cite{Qiu1998_ra1}. These bounds are asymptotically exact as $k \to 0$ and as $k \to 1$. The Qiu \emph{et al.} bounds for $k \in (0,1)$ are:

\begin{equation}
\ln\bigg( \frac{4}{\sqrt{1-k^2}} + (e^{\pi/2}-4)\sqrt{1-k^2}\bigg) < K(k) < \ln\bigg( \frac{4}{\sqrt{1-k^2}} + e^{\pi/2}-4\bigg).
\end{equation}

Fig.~\ref{fig_sharp_bound} (right) illustrates the sharp bounds for the complete elliptic integral $E(k)$. Yang and coauthors derived sharp lower and upper bounds for $E(k)$ in terms of elementary functions \cite{Yang2016_ra4,Yang2018_ra6}. The upper bound is also  exact in the two limiting cases. In the inset, the approximation $E(k)$ in this work intersects the upper bound at $k \approx 0.964$. This crossover suggests an interesting property that may merit further investigation within the framework of inequality analysis. The Yang \emph{et al.} bounds for $k \in (0,1)$ are of the forms:

\begin{equation}
\frac{7\pi}{22} \frac{\,\,1-(1-k^2)^{11/8}}{1-(1-k^2)^{7/8}} < E(k) < \frac{\,\,1-(1-k^2)^{\pi/(2\pi-4)}}{1-(1-k^2)^{1/(\pi-2)}}.
\end{equation}

As seen in the figure, the approximations in this work are highly competitive with sharp bounds from the pure mathematics literature, particularly for $K(k)$. Hence, the methods may also be of interest in the mathematical analysis of special functions.

\section{Discussions}

It can be argued that only the relatively simple forms are practical and useful. For if accuracy is the only concern for a particular application, exact value of the integrals can always be numerically evaluated, rendering any analytical form, no matter how accurate, irrelevant. It is in a situation when simplicity is as equally important, for example, in practical or prototyping engineering and architecture designs where extreme precision is not necessary, in a theoretical development that requires a tractable mathematical framework, and in a classroom where advanced mathematical topics are not accessible to students;  that an analytical form finds its usefulness.

It is instructive to discuss obvious applications of the proposed formulae to gauge if their performance would be suitable for a particular engineering design. The period of a nonlinear pendulum is proportional to the integral of the first kind \cite{Marion1995}. Data from Fig.~\ref{fig_percent} shows that the proposed $K(k)$ has a maximum error of 0.17\%. That means if the period is on the order of one second, the proposed $K(k)$ would produce an error no more than 1.7 milli-seconds. It is important to realize that this maximum error is not for a particular favorable condition, but for the full range of amplitudes. In addition, a circumference of an ellipse is proportional to the integral of the second kind. Data from Fig.~\ref{fig_percent} also shows that the proposed $E(k)$ has a maximum error of 0.03\%. That means if a circumference is on the order of one meter, the proposed $E(k)$ would produce an error no more than 0.3 milli-meters, for all elliptical conditions. In short, the formulae in this work provide roughly 1 in 1,000 accuracy for a typical engineering application.

The key finding is the ability to compute the integrals with exceptional accuracy on both limits of elliptical conditions. For example, the orbit of the Earth is almost circular with an eccentricity of only 0.0167 \cite{NASA_EarthFactSheet}. This is a favorable condition for the proposed $E(k)$. Knowing the semi-major axis of the Earth, one can compute the orbital path distance, which is the circumference of an ellipse. The proposed $E(k)$ would produce an error of only 7.4 centimeters, out of roughly 0.94 million kilometers that the Earth traverses around the Sun.

Aside from the obvious applications, they are also useful in analyzing exact solutions of a nonlinear pendulum and computing the exact inverse of $K$,
which are the subjects of the following discussions. 

\subsection{Nonlinear Pendulum}

Nonlinear pendulum has emerging applications in diverse fields of research, outside the realm of mathematical physics. For example, it can harvest energy from body movements or ocean waves \cite{Feng2024, Fu2023, Marszal2017}. Its solutions can teach a machine learning model to recognize nonlinear dynamics \cite{Krishnapriyan2023}. In addition, the phase space of an energy-conserving nonlinear pendulum from analytical solutions has not been easy traditionally. This is because traditional solutions $\theta(t)$ and $\omega(t)$ are in the forms of Jacobi elliptic functions, and so evaluating them required prior understanding of advanced mathematics, an expertise in narrowing fields of research. However, simple and exact solutions of nonlinear energy-conserving pendulum have recently been derived \cite{Chachiyo2005efd}. They are formally exact and written in terms of elementary functions, saving only one special, $K(k)$.

Therefore, the proposed $K(k)$ in this work is particularly useful when studying nonlinear behaviors of the system. Although an energy-conserving pendulum is nonlinear, it is not chaotic. Hence, an approximate form of $K(k)$ should not cause the solution to diverge from its exact trajectory, prompting meaningful interpretation both quantitatively and qualitatively. The phase space can be plotted with elementary functions, allowing simple access and broader collaboration among researchers from other fields of expertise.

\begin{figure}[h]
	\begin{center}
		\includegraphics[width=\textwidth]{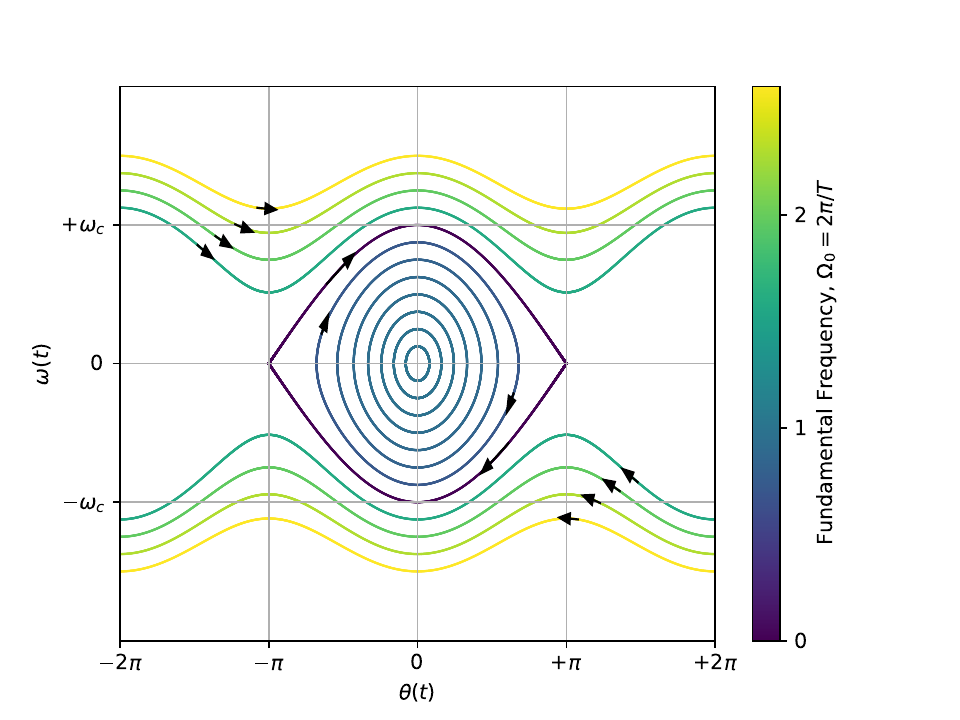}
		\caption{\label{fig_phase} Phase space using the exact solutions \cite{Chachiyo2005efd} and the proposed $K(k)$}
	\end{center}
\end{figure}

Fig.~\ref{fig_phase} shows the phase space of an energy-conserving nonlinear pendulum with physical parameters $\sqrt{L/g} = 1\,\text{second}$. The  $\theta(t)$ are as follows:

\begin{align}
	\omega_m < \omega_c: & \quad \theta(t) = \sum_{n\;\text{odd}} a_n \sin(n \Omega_0 t + n \delta), \label{eq_swinging_pendulum} \\
	\omega_m = \omega_c: & \quad \theta(t) = 2 \arcsin\big[ \tanh ( \sqrt{\tfrac{g}{L}} t + \delta) \, \big],  \\
	\omega_m > \omega_c: & \quad \theta(t) = \Omega_0 t + \delta  + \sum_{n \ge 1} b_n \sin(n \Omega_0 t + n \delta ). \label{eq_spinning_pendulum}
\end{align}

According to the exact frequency content solutions \cite{Chachiyo2005efd}, a trajectory is defined by its angular speed $\omega_m$ at the bottom. A set of 12 values of $\omega_m$ is chosen, resulting in 12 trajectories as shown in the graph. There are 3 distinctive patterns, corresponding to the 3 classes of motions: swinging, stopping, and spinning.

The swinging motion is represented by the 7 inner closed-loops. The smooth loops mean that their motions repeat with a period $T$. Since the motion is nonlinear, it is not a simple sinusoidal function, but rather a combination of infinite number of harmonics with fundamental angular frequency $\Omega_0 = 2\pi/T$. The loops are also color-coded with their fundamental frequencies. 

The stopping motion resembles an eye in the phase space with sharp edges on both sides. It is the $8^\text{th}$ trajectory with the darkest color. A motion that starts anywhere on the curve will approach either of the two sharp edges, $\theta = \pm \pi$, depending on its initial angular velocity. Physically, this is the class of motion where it starts with just enough kinetic energy to get to the unstable equilibrium peak, but can never quite get there. Due to conservation of energy, the closer it gets to the peak, the slower it moves. Thus, it takes forever to get there; and the period can be thought of as   infinity. The fundamental frequency is subsequently zero, as indicated by the darkest color of the trajectory.

The spinning motion are the 4 outer open paths. They may not appear periodic at first because the angle $\theta(t)$ keeps increasing as the pendulum spins in one direction. For example, the 4 trajectories at the top, starts with angle $- 2 \pi$ and continues to $-\pi,\,0,\,+\pi,\,+2\pi$. However, the motion is periodic in a sense that the angular velocity oscillates up and down, as clearly seen from the wavy feature in the phase diagram. Therefore,  the motion can still be written as a combination of harmonics, as precisely shown in Eq.~(\ref{eq_spinning_pendulum}).

Specifically, the period $T$ for both classes of motion are ($k = \sqrt{L/g}\,\omega_m/2$):

\begin{align}
\text{Swinging:} & \quad T = \sqrt{\frac{L}{g}} \, 4 \, K(k), \\
\text{Spinning:} & \quad T = \sqrt{\frac{L}{g}} \, \frac{2}{k} \, K(1/k).
\end{align}

The fact that exact solutions of nonlinear pendulum are often omitted in intermediate classical mechanics textbooks \cite{Marion1995, Taylor2004, Kibble2004} due to their mathematical complexity clearly shows that plotting the phase space using analytical solutions has always been difficult traditionally. The fact that it can now be done using only elementary functions such as sine and logarithm, also highlights the usefulness of the proposed formulae in this work.

\subsection{Never Failing Newton Initialization}

The approximate inverse of $K$ in Eq.~(\ref{eq_invK_thiswork}) can be used to find the \emph{exact} inverse using Newton root finding method. As mentioned earlier, an inverse of K is rarely implemented is a standard software package. In all likelihood, an in-house implementation is required. Newton root finding is a simple and powerful method for finding  an exact  mapping $K \rightarrow k$. Given an input $K$, an initial guess $k_0$ has to be given. The $k_n$ is then updated iteratively until the value does not change within a tolerant threshold. Specifically,

\begin{equation}
	k_{n+1} = k_n - \frac{K(k_n)-K}{K'(k_n)}.
\end{equation}

Here, $K$ is the input. $K(k_n)$ and $K'(k_n)$ are the complete elliptic integral and its derivative evaluated at $k_n$ respectively.

\begin{figure}[H]
	\begin{center}
		\includegraphics[width=\textwidth]{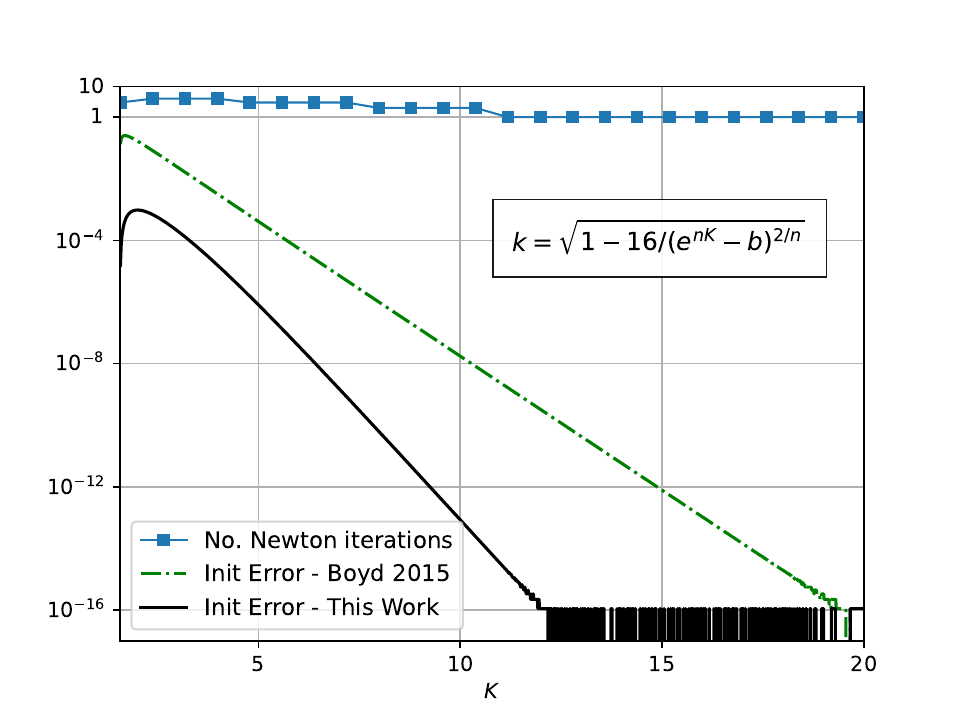}
		\caption{\label{fig_invK} Performance of the inverse of $K$ computation}
	\end{center}
\end{figure}

As discussed by Boyd \cite{Boyd2015}, for the root finding to be powerful, its initial value $k_0$ has to be Never Failing Newton Initialization (NFNI). Fig.~\ref{fig_invK} shows that the proposed inverse of $K$ in this work is indeed the case.

Fig.~\ref{fig_invK} shows the number of iterations (square markers) needed to compute the exact inverse of $K$ using Newton root finding method. The threshold is set to $10^{-14}$. From the graph, it takes less than 10 steps to complete; and  the number of steps shows downward trend. The range of $K$ tested is  $[\frac{\pi}{2}, 20]$ with $0.01$ interval. The computation is always successful for the entire range; evidently, the NFNI condition is met if Eq.~(\ref{eq_invK_thiswork}) is used as an initial guess.

The solid line is the error of the propose inverse as compared to the exact value calculated from the Newton root finding method. Maximum error of $0.12\%$ occurs at small $K$. Note that for $K > 12$ the error stays at $10^{-16}$, meaning the values are becoming numerically indistinguishable from the root finding method. Because Eq.~(\ref{eq_invK_thiswork}) is used as an initial guess, the error can also be interpreted as an error of the initial step, labeled \textquote{Init Error - This Work} in the graph.

The dash-dot line is the error of the initial step proposed by one of the pioneering works of Boyd \cite[Eq.43]{Boyd2015}. Both error profiles show similar qualitative features, peak at the lower range but rapidly decaying toward higher $K$ and most importantly NFNI. Quantitatively, the formula in this work exhibits better performance by 2 orders of magnitude at lower $K$ and roughly 6 orders at higher range. The difference in  performance is not totally significant if they are used as initial guesses because the Newton root finding should converge to the same value. However, for a simple prototyping task that 1 in 1,000 accuracy is acceptable, Eq.~(\ref{eq_invK_thiswork}) is a good standalone formula without the need of an iterative method.

\subsection{Exact Values using Arithmetic-Geometric Mean}

An illustration of AGM procedure for computing the exact $K(k), E(k)$ and their derivatives should be useful in a platform that special functions are not accessible such as web-based and firmware developments. With careful adaptation from the original recipe \cite[Eq.17.6.3-4]{Abramowitz1965}, the procedure in this work can compute $\{K, E, \frac{dK}{dk}, \frac{dE}{dk}\}$ simultaneously, with only one expensive square-root operation per iteration. See Supplemental Material for implementation in C and Python. The pseudo-code is shown in Algorithm \ref{alg_AGM}. 

\RestyleAlgo{ruled} 
\begin{algorithm}
\caption{Compute the exact integrals and their derivatives}\label{alg_AGM}
\DontPrintSemicolon
\SetKwFunction{FAGM}{AGM}

\FAGM{$k$}{
   	
  	\tcc*{special cases}
	\lIf{$k = 0$}{
		$K \gets \frac{\pi}{2}\;\quad E \gets \frac{\pi}{2} \quad dKdk \gets 0 \;\quad dEdk \gets 0$
	} 
	
	\lIf{$k = 1$}{
	$K \gets \infty \quad E \gets 1 \quad dKdk \gets \infty \quad \! dEdk \gets -\infty$\;
	}

	\If{$0 < k < 1$}{
		$a0 \gets 1$ \quad $b0 \gets \sqrt{1-k^2}$ \tcc*{K initialization}
		$tn \gets 1$ \quad $S \gets  k^2$ \tcc*{E initialization}
		\;
		\While{True}{
			
			$a \gets (a0+b0)/2$ \quad $b \gets \sqrt{a0 \, b0}$\;
			$tn \gets 2 \, tn  \qquad S \gets S + tn \, (a0-b0)^2/4$\;
			\;
			\lIf{$a = a0$}{break}
			$a0 \gets a$ \quad $b0 \gets b$
		}
		\;
		$K \gets \pi/(2 a)$ \quad $E \gets K(1-S/2)$\;
		$dKdk \gets (E-(1-k^2)K)/(k(1-k^2))$\;
		$dEdk \gets (E-K)/k$
	}
\;
\KwRet{K, E, dKdk, dEdk}
}
\end{algorithm}

The Newton root finding used in this work is shown in Algorithm \ref{alg_invK}. In this subroutine, the number of iteration is also returned for  performance analysis.

\begin{algorithm}
	\caption{Compute exact inverse of $K$}\label{alg_invK}
	\DontPrintSemicolon
	\SetKwFunction{FINV}{invK\_Newton}
	
	\FINV{$K$}{
		\;\;
		\tcc*{special case}
		\lIf{$K = \pi/2$}{$k \gets$ 0 \, nIter $\gets$ 0 \,\, \KwRet{k, \upshape nIter}}
		\;
		$n \gets$ $(\ln 4 - \ln \pi)/(\pi/2 - \ln 4)$ \tcc*{initial guess}
		$b \gets$ $\exp(n \pi /2) - 4^n$\;
		$k \gets$ $\sqrt{1-16/(e^{n K} - b)^{2/n}}$\;
		\;
		MAX = 100 \tcc*{safeguard}
		
		\For(\tcc*{main newton loop}){\upshape nIter = 1 $\to$ \upshape MAX}{
			$Kn,\,dKdk\_n \gets \text{AGM}(k)$\;
			$dk \gets -(Kn - K)/dKdk\_n$\;
			$k \gets k + dk$\;
			\lIf{$|dk| < 10^{-14}$}{break}
			}
		\;
		\KwRet{k, \upshape nIter}
	}
\end{algorithm}

\section{Theoretical Method}

Complete asymptotic behaviors of $K(k)$ and $E(k)$ are rarely available in mathematical handbooks. Fortunately, they are documented in the NIST Digital Library of Mathematical Functions (DLMF) \cite{NIST:DLMF} and serve as the foundation for the proposed formulae  in this work. The complete asymptotic behaviors are given in Ref.~\cite[Eq.19.12.1-2]{NIST:DLMF}:

\begin{align}
K(k) & = \sum_{m=0}^\infty \frac{{(\frac{1}{2})}_m {(\frac{1}{2})}_m}{m! m!} {k'}^{2m}\Big[ \ln(\frac{1}{k'}) + d(m)\Big], \\
E(k) & = 1 + \frac{1}{2}\sum_{m=0}^\infty \frac{{(\frac{1}{2})}_m {(\frac{3}{2})}_m}{{(2)}_m m!} {k'}^{2m+2}\Big[  \ln(\frac{1}{k'}) + d(m) - \frac{1}{(2m+1)(2m+2)}\Big].
\end{align}

\noindent where $k' = \sqrt{1-k^2}$. The parenthesis with subscript, for example ${(\frac{1}{2})}_m$, is  Pochhammer\textquoteright s symbol. The auxiliary function $d(m)$ has a recursive relation  $d(m+1) = d(m) - \frac{2}{(2m+1)(2m+2)}$, and $d(0) = \ln 4$. To find a balance between simplicity and accuracy, only the first term $m=0$ is kept. They become:

\begin{align}
	\text{Asymptotic limit:} \qquad K(k) & \approx \ln\Big[\frac{4}{\sqrt{1-k^2}}\Big],
	\label{eq_K_asymptotic} \\
	\text{Asymptotic limit:} \qquad E(k) & \approx 1 + \frac{1-k^2}{2}\ln \Big[ \frac{4/\!\sqrt{e}}{\sqrt{1-k^2}}  \Big]. \label{eq_E_asymptotic}
\end{align}

For the proposed $K(k)$ in Eq.~(\ref{eq_K_thiswork}), the form is motivated by inspecting the graph of asymptotic limit  for the full range of modulus $k$. As shown in Fig.~\ref{fig_KE}, the asymptotic curve still tracks the exact solution even though $k$ is small. The observation prompts an idea that a slight modification can be made to the asymptotic form such that the curve also agrees with the exact solution up to the second order Taylor expansion when $k \approx 0$, while maintaining its asymptotic behavior.

As $k \rightarrow 1$,  notice the divergent behavior of $4/\!\sqrt{1-k^2}$ in Eq.~(\ref{eq_K_asymptotic}). Because of this divergence, adding two small parameters will not disturb it, but will give us freedom to adjust the behavior at the other end, $k \rightarrow 0$.

The following form is proposed:

\begin{equation}
	K(k) \approx \frac{1}{n}\ln[ \Big(\frac{4}{\sqrt{1-k^2}}\Big)^n + b].
	\label{eq_K_form}
\end{equation}

The constant $b$ is such that $K(0) = \pi/2$ exactly. It can be shown that $n = \frac{\ln 4 - \ln \pi}{\pi/2 - \ln 4}$ and $b = e^{n\pi/2} - 4^n$ for the Eq.~(\ref{eq_K_form}) to meet the second order criteria (see Supplemental Material). As a result, however, the proposed form is surprisingly more accurate than the forth order Taylor expansion in the mid-modulus range as shown in Fig.~\ref{fig_percent}. 

Additionally, the form in Eq.~(\ref{eq_K_form}) is simple enough that its inverse can easily be  derived, as shown in Eq.~(\ref{eq_invK_thiswork}).

For the proposed $E(k)$ in Eq.~(\ref{eq_E_thiswork}), the same approximation technique can be used. First, locate the divergence in the asymptotic form. Second, add two small parameters without disturbing the divergence, while gaining more control of the behavior at the other limit, $k \rightarrow 0$. Third, evaluate parameters such that the Taylor expansion at the zero limit is satisfied.

The following form is proposed:

\begin{equation}
E(k)  \approx 1 + \frac{1-k^2}{2 n}\ln [ \Big( \frac{4/\!\sqrt{e}}{\sqrt{1-k^2}}\Big)^n  + b ].
	\label{eq_E_form}
\end{equation}

Second order Taylor expansion at $k=0$ is satisfied if $ n = \frac{  \ln(3\pi/2 - 4)\,}{ \ln 4 - \pi + 3/2 }$, and $b = e^{n(\pi-2)} - (4/\!\sqrt{e})^n$. Surprisingly again, the form is more accurate than forth order Taylor expansion in the mid-modulus range, as shown in Fig.~\ref{fig_percent}.

\section{Conclusion}
Simple closed-form formulae for the complete elliptic integral of the first and second kind are proposed. The inverse of the first kind is also presented. The formulae can be used in a multitude of physics and engineering applications with 1 in 1,000 accuracy for all elliptical conditions. The proposed formulae should be useful for practical or prototyping engineering and architecture designs. Especially the proposed inverse of $K$ can be used as a standalone approximation. If extreme accuracy is required, it is a suitable initial guess with Never Failing Newton Initialization property, which is an important step in the computation of the exact inverse.

\subsection*{Supplemental Material} A collection of pythons notebook,  python codes, and additional proofs is available at a GitHub repository,  \url{https://github.com/teepanis/elliptic-integrals}

\bibliographystyle{elsarticle-num}
\bibliography{manuscript_r2v1}

\begin{thebibliography}{10}
\expandafter\ifx\csname url\endcsname\relax
  \def\url#1{\texttt{#1}}\fi
\expandafter\ifx\csname urlprefix\endcsname\relax\def\urlprefix{URL }\fi
\expandafter\ifx\csname href\endcsname\relax
  \def\href#1#2{#2} \def\path#1{#1}\fi

\bibitem{Marion1995}
J.~Marion, S.~Thornton, Classical Dynamics of Particles and Systems, Saunders
  College Pub., 1995.

\bibitem{Feng2024}
W.~Feng, H.~Chen, Q.~Zou, D.~Wang, X.~Luo, C.~Cummins, C.~Zhang, S.~Yang,
  Y.~Su, A contactless coupled pendulum and piezoelectric wave energy
  harvester: Model and experiment, Energies 17 (2024) 876.
\newblock \href {https://doi.org/10.3390/en17040876}
  {\path{doi:10.3390/en17040876}}.

\bibitem{Fu2023}
H.~Fu, J.~Jiang, S.~Hu, J.~Rao, S.~Theodossiades, A multi-stable ultra-low
  frequency energy harvester using a nonlinear pendulum and piezoelectric
  transduction for self-powered sensing, Mechanical Systems and Signal
  Processing 189 (2023) 110034.
\newblock \href {https://doi.org/10.1016/j.ymssp.2022.110034}
  {\path{doi:10.1016/j.ymssp.2022.110034}}.

\bibitem{Marszal2017}
M.~Marszal, B.~Witkowski, K.~Jankowski, P.~Perlikowski, T.~Kapitaniak, Energy
  harvesting from pendulum oscillations, International Journal of Non-Linear
  Mechanics 94 (2017) 251--256.
\newblock \href {https://doi.org/10.1016/j.ijnonlinmec.2017.03.022}
  {\path{doi:10.1016/j.ijnonlinmec.2017.03.022}}.

\bibitem{Krishnapriyan2023}
A.~S. Krishnapriyan, A.~F. Queiruga, N.~B. Erichson, M.~W. Mahoney, Learning
  continuous models for continuous physics, Communications Physics 6 (2023)
  319.
\newblock \href {https://doi.org/10.1038/s42005-023-01433-4}
  {\path{doi:10.1038/s42005-023-01433-4}}.

\bibitem{Bair2002}
B.-W. Bair, Computer aided design of elliptical gears, Journal of Mechanical
  Design 124 (2002) 787--793.
\newblock \href {https://doi.org/10.1115/1.1485092}
  {\path{doi:10.1115/1.1485092}}.

\bibitem{Alom2016}
N.~Alom, S.~C. Kolaparthi, S.~C. Gadde, U.~K. Saha, Aerodynamic design
  optimization of elliptical-bladed savonius-style wind turbine by numerical
  simulations, in: Volume 6: Ocean Space Utilization; Ocean Renewable Energy,
  American Society of Mechanical Engineers, 2016.
\newblock \href {https://doi.org/10.1115/OMAE2016-55095}
  {\path{doi:10.1115/OMAE2016-55095}}.

\bibitem{Sutaji2021}
A.~Sutaji, M.~Yusvika, D.~D. D.~P. Tjahjana, S.~Hadi, Application of elliptical
  blade shape to enhance power generation of the savonius water turbine, IOP
  Conference Series: Materials Science and Engineering 1096 (2021) 012049.
\newblock \href {https://doi.org/10.1088/1757-899X/1096/1/012049}
  {\path{doi:10.1088/1757-899X/1096/1/012049}}.

\bibitem{Abramowitz1965}
M.~Abramowitz, I.~Stegun, Handbook of Mathematical Functions: With Formulas,
  Graphs, and Mathematical Tables, Applied mathematics series, Dover
  Publications, 1965.

\bibitem{Fukushima2013}
T.~Fukushima, Numerical computation of inverse complete elliptic integrals of
  first and second kinds, Journal of Computational and Applied Mathematics 249
  (2013) 37--50.
\newblock \href {https://doi.org/10.1016/j.cam.2013.02.003}
  {\path{doi:10.1016/j.cam.2013.02.003}}.

\bibitem{Boyd2015}
J.~P. Boyd, Four ways to compute the inverse of the complete elliptic integral
  of the first kind, Computer Physics Communications 196 (2015) 13--18.
\newblock \href {https://doi.org/10.1016/j.cpc.2015.05.006}
  {\path{doi:10.1016/j.cpc.2015.05.006}}.

\bibitem{Yang2020}
Z.-H. Yang, J.-F. Tian, Y.-R. Zhu, A rational approximation for the complete
  elliptic integral of the first kind, Mathematics 8 (2020) 635.
\newblock \href {https://doi.org/10.3390/math8040635}
  {\path{doi:10.3390/math8040635}}.

\bibitem{Zhu2022}
L.~Zhu, A natural approximation to the complete elliptic integral of the first
  kind, Mathematics 10 (2022) 1472.
\newblock \href {https://doi.org/10.3390/math10091472}
  {\path{doi:10.3390/math10091472}}.

\bibitem{Hinrichsen2021}
P.~F. Hinrichsen, Review of approximate equations for the pendulum period,
  European Journal of Physics 42 (2021) 015005.
\newblock \href {https://doi.org/10.1088/1361-6404/abad10}
  {\path{doi:10.1088/1361-6404/abad10}}.

\bibitem{Ganley1985}
W.~P. Ganley, Simple pendulum approximation, American Journal of Physics 53
  (1985) 73--76.
\newblock \href {https://doi.org/10.1119/1.13970} {\path{doi:10.1119/1.13970}}.

\bibitem{Lima2008}
F.~M.~S. Lima, Simple ‘log formulae’ for pendulum motion valid for any
  amplitude, European Journal of Physics 29 (2008) 1091--1098.
\newblock \href {https://doi.org/10.1088/0143-0807/29/5/021}
  {\path{doi:10.1088/0143-0807/29/5/021}}.

\bibitem{Qiu1998_ra1}
S.-L. Qiu, M.~K. Vamanamurthy, M.~Vuorinen, Some inequalities for the growth of
  elliptic integrals, SIAM Journal on Mathematical Analysis 29 (1998)
  1224--1237.
\newblock \href {https://doi.org/10.1137/S0036141096310491}
  {\path{doi:10.1137/S0036141096310491}}.

\bibitem{Alzer1998_ra2}
H.~Alzer, Sharp inequalities for the complete elliptic integral of the first
  kind, Mathematical Proceedings of the Cambridge Philosophical Society 124
  (1998) 309--314.
\newblock \href {https://doi.org/10.1017/S0305004198002692}
  {\path{doi:10.1017/S0305004198002692}}.

\bibitem{Alzer2004_ra3}
H.~Alzer, S.-L. Qiu, Monotonicity theorems and inequalities for the complete
  elliptic integrals, Journal of Computational and Applied Mathematics 172
  (2004) 289--312.
\newblock \href {https://doi.org/10.1016/j.cam.2004.02.009}
  {\path{doi:10.1016/j.cam.2004.02.009}}.

\bibitem{Yang2016_ra4}
Z.-H. Yang, Y.-M. Chu, W.~Zhang, Accurate approximations for the complete
  elliptic integral of the second kind, Journal of Mathematical Analysis and
  Applications 438 (2016) 875--888.
\newblock \href {https://doi.org/10.1016/j.jmaa.2016.02.035}
  {\path{doi:10.1016/j.jmaa.2016.02.035}}.

\bibitem{Yang2018_ra5}
Z.-H. Yang, W.-M. Qian, Y.~Chu, Monotonicity properties and bounds involving
  the complete elliptic integrals of the first kind, Mathematical Inequalities
  \& Applications (2018) 1185--1199\href
  {https://doi.org/10.7153/mia-2018-21-82} {\path{doi:10.7153/mia-2018-21-82}}.

\bibitem{Yang2018_ra6}
Z.-H. Yang, Sharp approximations for the complete elliptic integrals of the
  second kind by one-parameter means, Journal of Mathematical Analysis and
  Applications 467 (2018) 446--461.
\newblock \href {https://doi.org/10.1016/j.jmaa.2018.07.020}
  {\path{doi:10.1016/j.jmaa.2018.07.020}}.

\bibitem{Yang2020_ra7}
Z.-H. Yang, W.-M. Qian, en~Zhang, Y.~Chu, Notes on the complete elliptic
  integral of the first kind, Mathematical Inequalities \& Applications (2020)
  77--93\href {https://doi.org/10.7153/mia-2020-23-07}
  {\path{doi:10.7153/mia-2020-23-07}}.

\bibitem{Yang2021_ra8}
Z.-H. Yang, J.-F. Tian, Absolutely monotonic functions involving the complete
  elliptic integrals of the first kind with applications, Journal of
  Mathematical Inequalities (2021) 1299--1310\href
  {https://doi.org/10.7153/jmi-2021-15-87} {\path{doi:10.7153/jmi-2021-15-87}}.

\bibitem{NASA_EarthFactSheet}
{NASA National Space Science Data Center (NSSDCA)}, {Earth Fact Sheet},
  \url{https://nssdc.gsfc.nasa.gov/planetary/factsheet/earthfact.html},
  accessed: 2025-05-22 (2025).

\bibitem{Chachiyo2005efd}
T.~Chachiyo, \href{https://arxiv.org/abs/2504.16816}{Exact frequency-domain
  solutions for nonlinear pendulum-like dynamics}, arXiv:2504.16816 (2025).
\newblock \href {http://arxiv.org/abs/2504.16816} {\path{arXiv:2504.16816}}.
\newline\urlprefix\url{https://arxiv.org/abs/2504.16816}

\bibitem{Taylor2004}
J.~Taylor, Classical Mechanics, University Science Books, 2004.

\bibitem{Kibble2004}
T.~Kibble, F.~Berkshire, Classical Mechanics (5th Edition), World Scientific
  Publishing Company, 2004.

\bibitem{NIST:DLMF}
\href{https://dlmf.nist.gov/}{{\it NIST Digital Library of Mathematical
  Functions}}, \url{https://dlmf.nist.gov/}, Release 1.2.4 of 2025-03-15,
  f.~W.~J. Olver, A.~B. {Olde Daalhuis}, D.~W. Lozier, B.~I. Schneider, R.~F.
  Boisvert, C.~W. Clark, B.~R. Miller, B.~V. Saunders, H.~S. Cohl, and M.~A.
  McClain, eds.
\newline\urlprefix\url{https://dlmf.nist.gov/}

\end{thebibliography}

\newpage


\section*{Supplementary material}

# Supplemental Material

## Simple and accurate complete elliptic integrals for the full range of modulus

Teepanis Chachiyo teepanisc@nu.ac.th Department of Physics, Faculty of Science, Naresuan University, Phitsanulok 65000, Thailand.

Available at a GitHub repository: https://github.com/teepanis/elliptic-integrals

## Subroutines

Below are the subroutines introduced in this work.

### AGM(k):

It computes exact $(K, E, \frac{dK}{dk}, \frac{dE}{dk})$ to machine precision.

### K(k):

It computes the approximation for $K(k)$ proposed in this work.

### E(k):

It computes the approximation for $E(k)$ proposed in this work.

### invK(K):

It computes the approximation for an inverse of $K$ in this work.

### invK_Newton(K):

It computes an exact (up to $10^{-14}$ precision) inverse of $K$ using Newton root finding method and the proposed invK and initial guess. Additionally, nIter are returned. They are the number of Newton iterations used in the computation.

## Python Implementation

```
In [1]: import numpy as np

        # AGM K and E to machine precision from Abramowitz_Stegun Eq.17.6.3-4
        # If you want to check with scipy be aware that you need to use
        # scipy.special.ellipk(k**2) and scipy.special.ellipe(k**2)
        def AGM(k):

            # special cases
            if k==0:
                K = np.pi/2; E = np.pi/2; dKdk = 0; dEdk = 0
            if k==1:
                K = np.inf; E = 1; dKdk = np.inf; dEdk = -np.inf

            if 0 < k and k < 1:
                a0 = 1; b0 = np.sqrt(1-k**2)
                tn = 1; S  = k**2
                while True:
                    a  = (a0+b0)/2; b  = np.sqrt(a0*b0)
                    tn = 2*tn; S = S + tn*(a0-b0)**2/4

                    if a == a0: break
                    a0 = a; b0 = b

                K = np.pi/(2*a); E = K*(1-S/2)

                dKdk = (E-(1-k**2)*K)/(k*(1-k**2))
                dEdk = (E-K)/k

            return K, E, dKdk, dEdk
```

```python
#
# simple and accurate series
#

def K(k):
    #n = (np.log(4) - np.log(np.pi))/(np.pi/2 - np.log(4))
    #b = np.exp(n*np.pi/2) - 4**n
    # -- pre-computed for speed ---
    n = 1.3092785997521463
    b = 1.678061276031407
    
    # avoid numerical divide-by-zero
    if k==1: return np.inf
    
    return 1/n*np.log( (4/np.sqrt(1-k**2))**n + b)

def E(k):
    #n = np.log(3*np.pi/2 - 4)/(np.log(4) - np.pi + 3/2)
    #b = np.exp(n*(np.pi-2)) - (4/np.sqrt(np.e))**n
    
    # --- pre-computed for speed ---
    n = 1.3283723627880784
    b = 1.3103755722411727
    
    # avoid numerical divide-by-zero
    if k==1: return 1
    
    return 1 + 1/2*(1-k**2)/n*np.log( (4/np.sqrt(np.e)/np.sqrt(1-k**2))**n + b  )

def invK(K):
    #n = (np.log(4) - np.log(np.pi))/(np.pi/2 - np.log(4))
    #b = np.exp(n*np.pi/2) - 4**n
    # -- pre-computed for speed ---
    n = 1.3092785997521463
    b = 1.678061276031407
    
    # avoid numerical divide-by-zero
    if K==np.pi/2: return 0
    
    return np.sqrt(1-16/(np.exp(n*K)-b)**(2/n))

def invK_Newton(K):
    
    # special case
    if K==np.pi/2: return 0,0
    
    # initial guess
    k = invK(K)
    
    # safeguard
    MAX=100
    
    for nIter in range(1,MAX):
        Kn,En,dKdk_n,dEdk_n = AGM(k)
        dk = -(Kn-K)/dKdk_n
        k = k + dk
        if np.abs(dk)<1e-14: break
    
    return k, nIter
```

## C Implementation

```c
//
// Compute the exact elliptic integrals and their derivatives
//
void AGM(
double k, // input
double *K, double *E, double *dKdk, double *dEdk){ // output

    double a,b,a0,b0,tn,S;

    // special cases
    if(k==0){ *K = M_PI/2; *E = M_PI/2;
              *dKdk = 0; *dEdk = 0; }
    if(k==1){ *K = INFINITY; *E = 1;
              *dKdk = INFINITY; *dEdk = -INFINITY; }

    if(0 < k && k < 1){
        a0 = 1; b0 = sqrt(1-k*k);
        tn = 1; S  = k*k;
        while(1){
            a = (a0+b0)/2; b = sqrt(a0*b0);
            tn = 2*tn; S = S + tn*(a0-b0)*(a0-b0)/4;
            if(a0==a) break;
            a0 = a; b0 = b;
        }
        *K = M_PI/(2*a); *E = *K *(1-S/2);
        *dKdk = (*E-(1-k*k)*(*K))/(k*(1-k*k));
        *dEdk = ((*E)-(*K))/k;
    }
}

//
// simple and accurate series
//
double K(double k){
    //double n = (log(4) - log(M_PI))/(M_PI/2 - log(4));
    //double b = exp(n*M_PI/2) - pow(4,n);
    // --- pre-computed for speed ---
    double n = 1.3092785997521463;
    double b = 1.678061276031407;

    // avoid numerical divide-by-zero
    if(k==1) return INFINITY;

    return 1/n*log(pow(4/sqrt(1-k*k),n)+b);
}

double invK(double K){
    //double n = (log(4) - log(M_PI))/(M_PI/2 - log(4));
    //double b = exp(n*M_PI/2) - pow(4,n);
    // --- pre-computed for speed ---
    double n = 1.3092785997521463;
    double b = 1.678061276031407;

    // avoid numerical divide-by-zero
    if(K==M_PI/2) return 0;

    return sqrt(1-16/pow(exp(n*K)-b,2/n));
}
```

```c
double E(double k){
    //double n = log(3*M_PI/2 - 4)/(log(4) - M_PI + 3./2);
    //double b = exp(n*(M_PI-2)) - pow(4/sqrt(M_E),n);
    // --- pre-computed for speed ---
    double n = 1.3283723627880784;
    double b = 1.3103755722411727;

    // avoid numerical divide-by-zero
    if(k==1) return 1;

    return 1 + 1./2*(1-k*k)/n*log(
        pow(4/sqrt(M_E)/sqrt(1-k*k),n) + b );
}

//
// Compute exact inverse of K
//
void invK_Newton(
double K, // input
double *k, int *nIter){ // output

    double dk, Kn, En, dKdk_n, dEdk_n;
    int MAX=100;

    // special case
    if(K==M_PI/2){ *k = 0.0; return; }

    // initial guess
    *k = invK(K);

    for((*nIter)=1; (*nIter)<MAX; (*nIter)++){
        AGM(*k, &Kn, &En, &dKdk_n, &dEdk_n);
        dk = -(Kn-K)/dKdk_n;
        *k = *k + dk;
        if(fabs(dk)<1e-14) break;
    }
}
```

# Results

## Figure 1: Exact K(k) and E(k)

In [2]:
```python
import numpy as np
import matplotlib.pyplot as plt

# allocate amplitude thetaM and x
thetaM = np.arange(0,np.pi,.004)
k = np.sin(thetaM/2)
N = len(k)

# compute exact K(k) and E(k)
K_exact = np.zeros(N)
E_exact = np.zeros(N)
for i in range(0,N):
    K_exact[i], E_exact[i],dKdk,dEdk = AGM(k[i])


# allocate array for comparisons
K_asym = np.zeros(N)
K_t4 = np.zeros(N)
E_t4 = np.zeros(N)
K_ThisWork = np.zeros(N)
E_ThisWork = np.zeros(N)
K_lima2008 = np.zeros(N)
K_ganley1985 = np.zeros(N)

for i in range(N):

    # (Lima 2008)
    if k[i] > 0.0:
        a = np.sqrt(1-k[i]**2)
        K_lima2008[i] = np.pi/2*( a**2/(1-a)*np.log(1/a) +  2*k[i]**2/np.pi*np.log(4/a)   )
    else:
        K_lima2008[i] = np.pi/2

    K_asym[i] = np.log(4/np.sqrt(1-k[i]**2))
    K_t4[i] = np.pi/2*(1 + 1/4*k[i]**2 + ((1*3)/(2*4))**2*k[i]**4 )
    E_t4[i] = np.pi/2*(1 - 1/4*k[i]**2 - ((1*3)/(2*4))**2*k[i]**4/3 )

    # (Ganley 1985)
    alpha = np.sqrt(3)/2*thetaM[i]
    if alpha > 0.0: K_ganley1985[i] = np.pi/2*np.sqrt(alpha/np.sin(alpha))
    else: K_ganley1985[i] = 1.0

    n = (np.log(4) - np.log(np.pi))/(np.pi/2 - np.log(4))
    b = np.exp(n*np.pi/2) - 4**n
    K_ThisWork[i] = K(k[i])
    E_ThisWork[i] = E(k[i])

plt.plot(k, K_exact,'k.',markevery=20, label='K -- Exact')
plt.plot(k, E_exact,'bo',markevery=40, markersize=5, markerfacecolor='none', label='E -- Exact')
plt.plot(k, K_asym,'g-.', label='K -- Asymptotic')
#plt.plot(k, K_t4,'g--', label='K -- 4th Order Taylor')
plt.plot(k, K_ThisWork, 'r-',linewidth=1,label='K -- This Work')
plt.plot(k, E_ThisWork, 'b',linewidth=1,label='E -- This Work',
         linestyle=(0, (3, 1, 1, 1)))

plt.text(30/180,2.969/4*np.pi,'$K(k) \\approx  \\dfrac{1}{n} '
         '\\ln [   \\left(\\frac{4}{\\sqrt{1-k^2}}\\right)^n + b ]$', fontsize=12, \
         bbox=dict(boxstyle='square,pad=0.8', facecolor='white', alpha=0.9, edgecolor='black', linewidt

#plt.title('Complete Elliptical Integral of the First Kind')
plt.xlabel('$k$')
plt.ylabel('$K(k), E(k)$')
plt.yticks([0,np.pi/2,np.pi,3*np.pi/2,2*np.pi],['0','$\pi/2$','$\pi$','$3\\pi/2$','$2\pi$'])
plt.ylim([np.pi/4,5*np.pi/4])
plt.xlim([0,1])
plt.legend()
plt.grid()
#plt.savefig('fig_exact_KE.pdf')
plt.show()
```

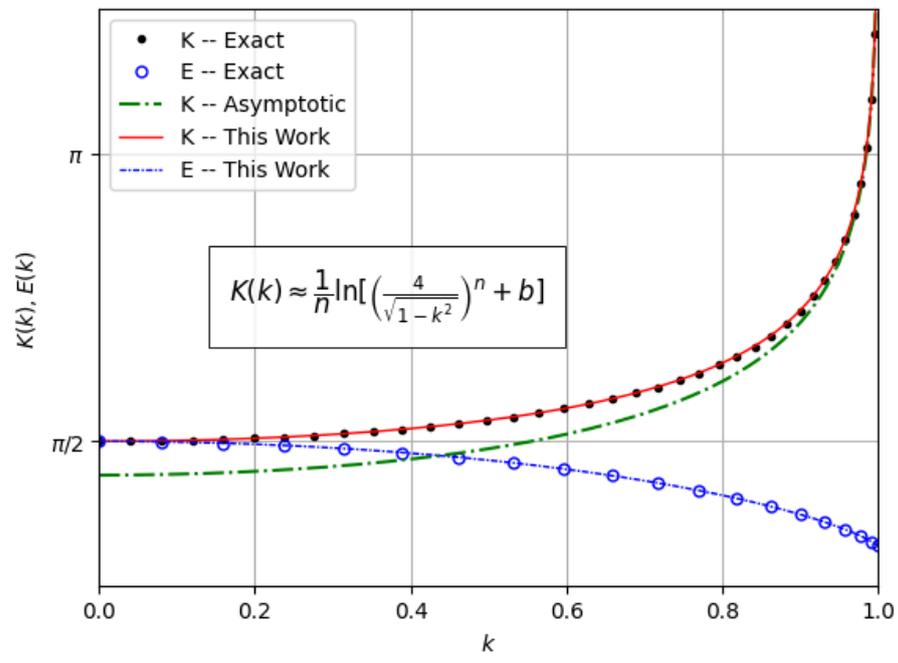

## Figure 2: Comparisons to Existing Analytical Forms

```python
In [3]:  #
         # Error comparison
         #
         fig, ax = plt.subplots()

         ax.plot(thetaM*180/np.pi, np.abs(K_ganley1985-K_exact)*100./K_exact,':', label='K -- Ganley 1985')
         ax.plot(thetaM*180/np.pi, np.abs(K_lima2008-K_exact)*100./K_exact,'-.', label='K -- Lima 2008')

         ax.plot(thetaM*180/np.pi, np.abs(K_t4-K_exact)*100./K_exact,'g--', label='K -- 4th Order Taylor')
         ax.plot(thetaM*180/np.pi, np.abs(E_t4-E_exact)*100./E_exact,'k--', label='E -- 4th Order Taylor')

         ax.plot(thetaM*180/np.pi, np.abs(K_ThisWork-K_exact)*100./K_exact,'r-',label='K -- This Work')
         ax.plot(thetaM*180/np.pi, np.abs(E_ThisWork-E_exact)*100./K_exact,'b',label='E -- This Work',
                 linestyle=(0, (3, 1, 1, 1)))

         #plt.title('Error compared to the exact K(k) and E(k)')
         plt.xlabel('$\\theta_{max}\;$ [degrees]')
         plt.ylabel('% |Error|')
         plt.xlim([0,180])
         plt.ylim([1e-8,1e2])
         secax = ax.secondary_xaxis('top')
         secax.set_xticks([0,180],['0','1'])
         secax.set_xlabel('modulus $k$')
         plt.legend()
         plt.grid(axis='y')
         plt.xticks([0,30,60,90,120,150,180])
         plt.yscale('log')

         plt.show()
         #plt.savefig('fig_percent_error.pdf')

         ## K ##
         print('--- K ---')
         print('Maximum % |Error|')
         #print('Lima 2008: %.2f' % ( max(np.abs(K_lima2008-K_exact)*100./K_exact) ) )
         print('This Work K(k): %.4f' % ( max(np.abs(K_ThisWork-K_exact)*100./K_exact) ) )

         print('\nAverage % |Error|')
         #print('Lima 2008: %.2f' % ( np.average(np.abs(K_lima2008-K_exact)*100./K_exact) ) )
         print('This Work K(k): %.4f' % ( np.average(np.abs(K_ThisWork-K_exact)*100./K_exact) ) )

         ## E ##
         print('\n--- E ---')
         print('Maximum % |Error|')
         print('This Work E(k): %.4f' % ( max(np.abs(E_ThisWork-E_exact)*100./E_exact) ) )
         print('\nAverage % |Error|')
         print('This Work E(k): %.4f' % ( np.average(np.abs(E_ThisWork-E_exact)*100./E_exact) ) )
```

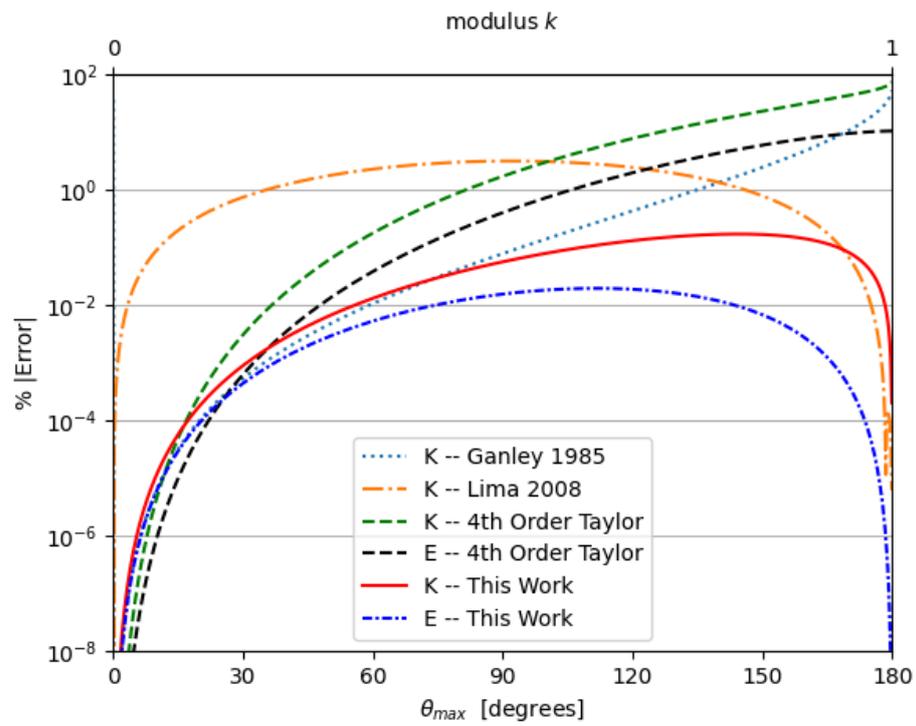

```
--- K ---
Maximum % |Error|
This Work K(k): 0.1699

Average % |Error|
This Work K(k): 0.0648

--- E ---
Maximum % |Error|
This Work E(k): 0.0333

Average % |Error|
This Work E(k): 0.0130
```

# Theoretical Method

## Derivation of the Constant $n$ and $b$ for K(k)

Below is a python code that symbollically derivatives of the proposed form, evaluated at $k=0$. The derivatives can then be compared to that of the exact $K(k)$. Demanding that they agree upto the 2nd order will ensure that the proposed form approaches the exact limit when the amplitudes is small.

In [4]:
```python
## checking taylor expansion of the proposed form
import sympy as sym

kk = sym.symbols('k', real=True)
bb = sym.symbols('b', real=True)
nn = sym.symbols('n', real=True)

bb = sym.exp(nn*sym.pi/2) - 4**nn

# the proposed form
K0 = 1/nn*sym.ln((4/sym.sqrt(1-kk**2))**nn + bb)

from IPython.display import display
print('0th Derivative: '); display(K0.subs(kk,0))
K1 = K0.diff(kk)
print('1st Derivative: '); display(K1.subs(kk,0))
K2 = K1.diff(kk)
print('2nd Derivative: '); display(K2.subs(kk,0))
K3 = K2.diff(kk)
print('3rd Derivative: '); display(K3.subs(kk,0))
```

0th Derivative:
$$\frac{\pi}{2}$$
1st Derivative:
$$0$$
2nd Derivative:
$$4^n e^{-\frac{\pi n}{2}}$$
3rd Derivative:
$$0$$

The value $b$ such that $K(0) = \frac{\pi}{2}$ exactly, is quite easy to find. In this case,

$$b = e^{n\pi/2} - 4^n \tag{S1}$$

Taylor series (sometimes called power series) of the exact $K(k)$ is

$$K(k) = \frac{\pi}{2}[1 + (\frac{1}{2})^2 k^2 + (\frac{1 \cdot 3}{2 \cdot 4})^2 k^4 + (\frac{1 \cdot 3 \cdot 5}{2 \cdot 4 \cdot 6})^2 k^6 + \cdots] \tag{S2}$$

Therefore, $K(0) = \frac{\pi}{2}$ and $K''(0) = \frac{\pi}{4}$. From the output of the python code above, it follows that

$$4^n e^{-\frac{2\pi}{2}} = \frac{\pi}{4} \tag{S3}$$

which lead to

$$n = \frac{\ln 4 - \ln \pi}{\pi/2 - \ln 4} \quad \text{and} \quad b = e^{n\pi/2} - 4^n \tag{S4}$$

# Derivation of the Constant $n$ and $b$ for E(k)

```
In [5]:  ## checking taylor expansion of the proposed form
         import sympy as sym

         kk = sym.symbols('k', real=True)
         bb = sym.symbols('b', real=True)
         nn = sym.symbols('n', real=True)

         bb = sym.exp(nn*(sym.pi-2)) - (4/sym.sqrt(sym.E))**nn

         # the proposed form
         K0 = 1 + (1-kk**2)/2/nn*sym.ln((4/sym.sqrt(sym.E)/sym.sqrt(1-kk**2))**nn + bb)

         from IPython.display import display
         print('0th Derivative: '); display(K0.subs(kk,0))
         K1 = K0.diff(kk)
         print('1st Derivative: '); display(K1.subs(kk,0))
         K2 = K1.diff(kk)
         print('2nd Derivative: '); display(K2.subs(kk,0))
         K3 = K2.diff(kk)
         print('3rd Derivative: '); display(K3.subs(kk,0))
```

```
0th Derivative:
```
$$\frac{\pi}{2}$$
```
1st Derivative:
0
2nd Derivative:
```
$$\frac{\left(\frac{4}{e^{\frac{1}{2}}}\right)^{n} e^{-n(-2+\pi)}}{2} - \pi + 2$$
```
3rd Derivative:
0
```

The value $b$ such that $E(0) = \frac{\pi}{2}$ exactly, is quite easy to find. In this case,

$$b = e^{n(\pi-2)} - (4/\sqrt{e})^n \tag{S5}$$

Taylor series (sometimes called power series) of the exact $E(k)$ is

$$E(k) = \frac{\pi}{2}[1 - (\frac{1}{2})^2 \frac{k^2}{1} - (\frac{1\cdot 3}{2\cdot 4})^2 \frac{k^4}{3} - (\frac{1\cdot 3\cdot 5}{2\cdot 4\cdot 6})^2 \frac{k^6}{5} - \cdots] \tag{S6}$$

Therefore, $E(0) = \frac{\pi}{2}$ and $E''(0) = -\frac{\pi}{4}$. From the output of the python code above, it follows that

$$\frac{(4/\sqrt{e})^n e^{n(2-\pi)}}{2} - \pi + 2 = -\frac{\pi}{4} \tag{S7}$$

which lead to

$$n = \frac{\ln(3\pi/2 - 4)}{\ln 4 - \pi + 3/2} \quad \text{and} \quad b = e^{n(\pi-2)} - (4/\sqrt{e})^n \tag{S8}$$

# Discussions

## Earth Orbital Distance

In [6]:
```python
# Info from Nasa Earth fact-sheet
R = 149.598e6
eccen = .0167

K_exact,E_exact,dKdk,dEdk = AGM(eccen)
print("Assume Circle:      [km]", 2*np.pi*R/1e3)
print("Ellipse: Exact      [km]", 4*R*E_exact/1e3)
print("Ellipse: This Work  [km]", 4*R*E(eccen)/1e3)
print("Error:              [cm]", 4*R*(E(eccen)-E_exact)*1e2)
```

```
Assume Circle:      [km] 939951.9555834518
Ellipse: Exact      [km] 939886.4163558404
Ellipse: This Work  [km] 939886.416429983
Error:              [cm] 7.414262979565933
```

## Figure 3 Phase-Space

In [7]:
```python
import numpy as np
import matplotlib.pyplot as plt

##
## the Chiang Khruea solutions (CKS)
##
## plotting phase-space
def phase_space_CKS(omegaM, g, L, maxN=20):
    
    # trivial case
    if omegaM == 0:
        theta = np.zeros(1)
        omega = np.zeros(1)
        Omega = np.zeros(1)
        return theta, omega, Omega
    
    omegac = 2*np.sqrt(g/L)
    
    #
    # compute time sampling
    #
    dt = 0.005
    
    # swinging
    if omegaM < omegac:
        k = np.sqrt(L/g)*omegaM/2
        T = 4*np.sqrt(L/g)*K(k)
        delta = -np.pi/2
        t = np.arange(0,T/2,dt)
        Omega = np.ones(len(t))*2*np.pi/T
    
    # stopping
    elif omegaM == omegac:
        # start with 99.9% of -PI
        theta0 = -99.9/100*np.pi
        delta = 1/2*np.sqrt(L/g)*np.log((1+np.sin(theta0/2))/(1-np.sin(theta0/2)))
        # time to reach 99.9% of +PI
        T = ( np.arctanh(np.sin(99.9/100*np.pi/2)) )/np.sqrt(g/L) -  delta
        t = np.arange(0,T,dt)
        Omega = np.zeros(len(t))
    
    # spinning
    else:
        k = np.sqrt(L/g)*omegaM/2
        T = 2*np.sqrt(L/g)/k*K(1/k)
        delta = -2*np.pi
        t = np.arange(0,2*T,dt)
        Omega = np.ones(len(t))*2*np.pi/T
    
    
    theta = np.zeros(len(t))
    omega = np.zeros(len(t))
    
    #
    # plot upper-half phase space
    #
    
    # swinging motion
    if omegaM < omegac:
        
        # compute coefficients
        a = np.zeros(maxN)
        kappa = K(np.sqrt(1-k**2))/K(k)
        for n in range(1,maxN,2):
            a[n] = 4/n/np.cosh(kappa*n*np.pi/2)
        
        # sum contributions at time t
        for i in range(len(t)):
            st = 0
            so = 0
            for n in range(0,maxN):
                st = st + a[n]*np.sin(2*n*np.pi/T*t[i] + n*delta)
                so = so + 2*n*np.pi/T*a[n]*np.cos(2*n*np.pi/T*t[i] + n*delta)
            theta[i] = st
            omega[i] = so
```

```python
        # stopping motion
        elif omegaM == omegac:
            for i in range(len(t)):
                theta[i] = 2*np.arcsin( np.tanh(np.sqrt(g/L)*(t[i]+delta)  ) )
                omega[i] = 2/np.sqrt( 1 - np.tanh(np.sqrt(g/L)*(t[i]+delta))**2 ) \
                            /np.cosh( np.sqrt(g/L)*(t[i]+delta)  )**2                 \
                            *np.sqrt(g/L)

        # spinning motion
        else:

            # compute coefficients
            b = np.zeros(maxN)
            kappa = K(np.sqrt(1-1/k**2))/K(1/k)
            for n in range(1,maxN):
                b[n] = 2/n/np.cosh(n*np.pi*kappa)

            # sum contributions at time t +
            for i in range(len(t)):
                st = 0
                so = 0
                for n in range(0,maxN):
                    st = st + b[n]*np.sin(2*n*np.pi/T*t[i] + n*delta)
                    so = so + 2*n*np.pi/T*b[n]*np.cos(2*n*np.pi/T*t[i] + n*delta)
                theta[i] = 2*np.pi/T*t[i] + delta + st
                omega[i] = 2*np.pi/T + so

    return theta, omega, Omega
```

```python
m = 1.0
g = 9.8
# linear (natural) angular frequency of a pendulum
OmegaL = 1
L = g/OmegaL**2

omegac = 2*np.sqrt(g/L)
omegaM = np.linspace(0,1.5,13)*omegac

# delete the first with zero speed
omegaM = np.delete(omegaM,0)

theta = np.zeros(0)
omega = np.zeros(0)
Omega = np.zeros(0)
for i in range(len(omegaM)):
    thetaT, omegaT, OmegaT = phase_space_CKS(omegaM[i], g, L)

    theta = np.append(theta, thetaT)
    omega = np.append(omega, omegaT)
    Omega = np.append(Omega, OmegaT)

    # draw customized arrows
    def draw_arrow(sI,fI):
        plt.arrow(thetaT[sI], omegaT[sI],
                  thetaT[fI]-thetaT[sI], omegaT[fI]-omegaT[sI],
                  shape='full', lw=1, length_includes_head=False, head_width=.15,facecolor='k',zorder=
        # mirror
        plt.arrow(-thetaT[sI], -omegaT[sI],
           -thetaT[fI]+thetaT[sI], -omegaT[fI]+omegaT[sI],
              shape='full', lw=1, length_includes_head=False, head_width=.15,facecolor='k',zorder=3)

    if i==6: draw_arrow(100,100+15);
    if i==7: draw_arrow(1200,1200+60);
    if i>=8: draw_arrow(i*20,i*20+15);


## upper-half
plt.scatter(theta,omega,c=Omega,cmap='viridis',s=1,marker='o',linewidths=.1)

## mirror lower-half
plt.scatter(-theta,-omega,c=Omega,cmap='viridis',s=1,marker='o',linewidths=.1)

plt.xlim(-2*np.pi,2*np.pi)
plt.ylim(-2*omegac,2*omegac)
plt.xticks([-2*np.pi,-np.pi,0,np.pi,2*np.pi],
           ['$-2\pi$','$-\pi$','$0$','$+\pi$','$+2\pi$'])
plt.yticks([-omegac,0,+omegac],
```

```
              ['$-\\omega_c$','$0$','$+\\omega_c$'])
plt.xlabel('$\\theta(t)$')
plt.ylabel('$\\omega(t)$')

plt.colorbar(label='Fundamental Frequency, $\\Omega_0 = 2\pi/T$',ticks=[0,1,2])
plt.grid()
plt.show()
#plt.savefig('fig_phase_space.pdf')
```

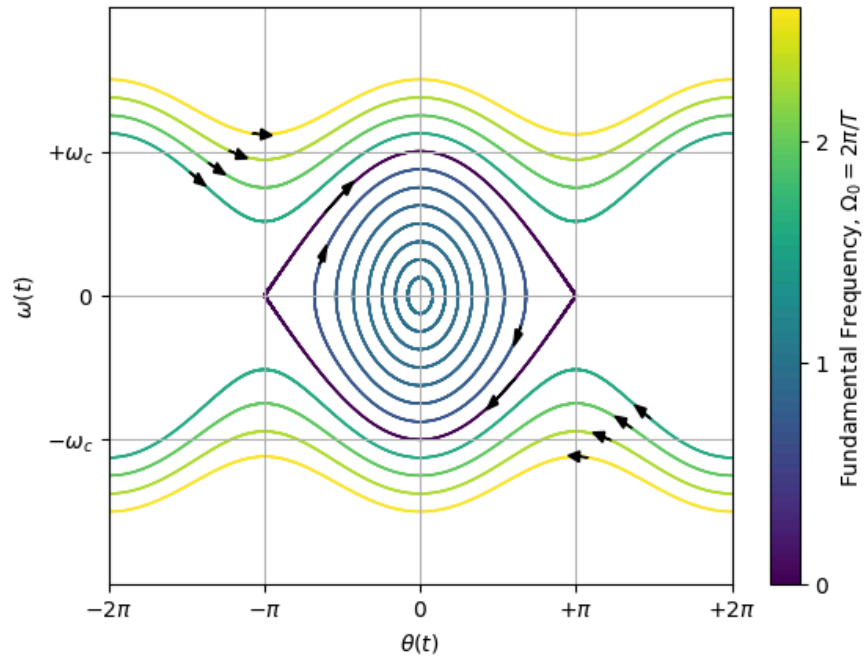

## Figure 4 Inverse K performance

```python
In [9]: import numpy as np
import matplotlib.pyplot as plt

dK = 0.01
K = np.arange(np.pi/2+dK,20,dK)
k_exact = np.zeros(len(K))
k_thiswork = np.zeros(len(K))
k_Boyd = np.zeros(len(K))
nIter = np.zeros(len(K))

for i in range(len(K)):
    k_exact[i], nIter[i] = invK_Newton(K[i])
    k_thiswork[i] = invK(K[i])
    
    d0 = 8/np.pi
    d1 = 4*(4-np.pi)/np.pi/(np.pi-4*np.log(2))
    d2 = d1/2
    lambdaB = K[i]-np.pi/2
    k_Boyd[i] = 1-np.exp(-lambdaB*(d0+d1*lambdaB)/(1+d2*lambdaB))

plt.plot(K[::80], nIter[::80],'s-',label='No. Newton iterations',
         markersize=5,linewidth=1)
plt.plot(K, np.abs(k_Boyd-k_exact),'g-.',label='Init Error - Boyd 2015')
plt.plot(K, np.abs(k_thiswork-k_exact),'k-',label='Init Error - This Work')
plt.yscale('log')
#plt.title('Performance of inverse K computation')
plt.xticks([0,5,10,15,20])
plt.yticks([1e-16,1e-12,1e-8,1e-4,1e0,1e1],
           ['$10^{-16}$','$10^{-12}$','$10^{-8}$','$10^{-4}$','$1$','$10$'])
plt.ylim(1e-17,1e1)
plt.xlim(np.pi/2,20)
plt.xlabel('$K$')
plt.legend(fontsize=11,loc='lower left')
plt.grid()

plt.text(11.3,4.25e-5,'$k = \\sqrt{1-16/(e^{nK}-b)^{2/n}}$', fontsize=12, \
         bbox=dict(boxstyle='square,pad=0.8', facecolor='white',
                   alpha=0.9, edgecolor='black', linewidth=0.7))

#plt.savefig('fig_performance_inverseK.pdf')
plt.show()

print('Max Error: ', np.max(np.abs(k_thiswork-k_exact)))
print('Max Percent Error: ', np.max(np.abs((k_thiswork-k_exact)*100./k_exact)))
```

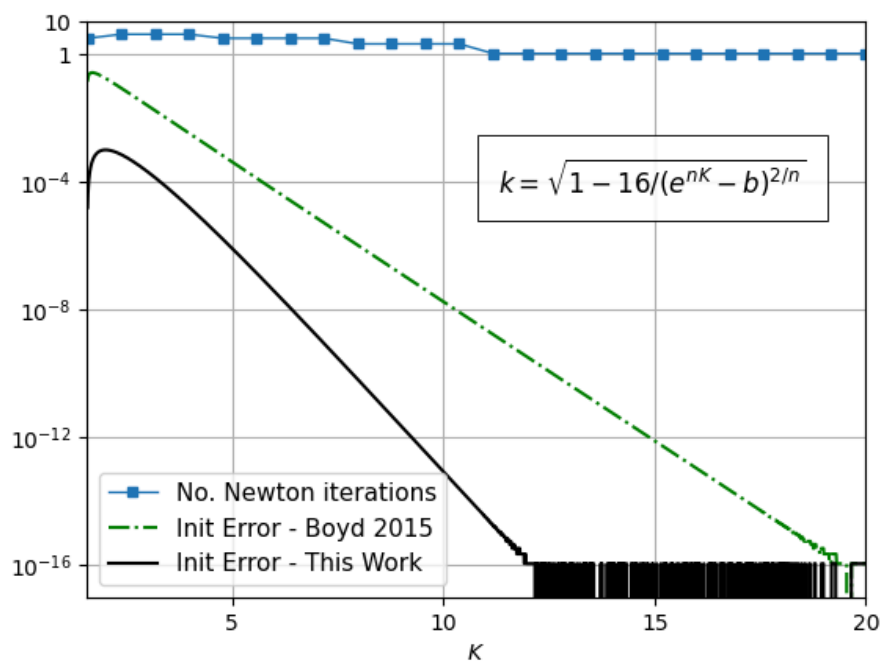

```
Max Error:  0.0009696795924981627
Max Percent Error:  0.12410178822665986
```



\end{document}